\documentclass[12pt]{article}
\usepackage[english]{babel}
\usepackage{graphicx,epsfig}
\makeindex \textwidth 170mm \textheight 220mm \topmargin -5mm
\newcommand{\be}{\begin{equation}}
\newcommand{\ee}{\end{equation}}
\newcommand{\bea}{\begin{eqnarray}}
\newcommand{\eea}{\end{eqnarray}}

\def\somma#1#2{\sum_{\rm #1}^{\rm #2}}

\def\rfr#1{eq. (\ref{#1})}

\def\Rfr#1{Eq. (\ref{#1})}

\def\bb{\bibitem}
\def\eqi{\begin{equation}}
\def\eqf{\end{equation}}
\def\eqia{\begin{eqnarray}}
\def\eqfa{\end{eqnarray}}

\def\rp#1#2{{#1\over#2}}

\def\lb#1{\label{#1}}
\def\lg{LAGEOS}
\def\lgg{LAGEOS II}
\def\lt{Lense-Thirring\ }
\def\grc{gravitomagnetic}

\def\sat{satellite}

\def\b{\beta}
\def\ga{\gamma}
\def\d{\delta}

\def\et{\eta}

\def\k{\kappa}

\def\n{\nu}

\def\og{\omega}

\def\D{\Delta}

\def\O{\Omega}

\def\ppn{\rp{2+2\ga-\b}{3}}

\begin{document}
\begin{titlepage}
\begin{flushright}
\today\\
BARI-TH/00\\
\end{flushright}
\vspace{.5cm}
\begin{center}
{\LARGE Measuring the Relativistic Perigee Advance with Satellite
Laser Ranging} \vspace{.5cm}
\quad\\
{Lorenzo Iorio$^{\dag}$, Ignazio Ciufolini$^{\ddag}$, Erricos C. Pavlis$^{*}$\\
\vspace{0.3cm}
\quad\\
{\dag}Dipartimento di Fisica dell' Universit{\`{a}} di Bari, via
Amendola 173, 70126, Bari, Italy\\ \vspace{0.5cm}
\quad\\
{\ddag}Dipartimento di Ingegneria dell'Innovazione
dell'Universit{\`{a}} di Lecce, via per Arnesano, 73100, Lecce,
Italy
\\
\vspace{0.3cm}
\quad\\
{*} Joint Center for Earth Systems Technology, JCET/UMBC and NASA
Goddard Space Flight Center, Greenbelt, Maryland, 20771-0001
U.S.A.}\\ \vspace{.5cm}

{\bf Abstract\\}
\end{center}
{\noindent \footnotesize The pericenter advance of a test body by
a central mass is one of the classical tests of General
Relativity. To day this effect is measured with radar ranging by
the perihelion shift of Mercury and other planets, in the
gravitational field of Sun, with a relative accuracy of the order
of $10^{\rm -2}$--$10^{\rm -3}$. In this paper we explore the
possibility of a measurement of the pericenter advance in the
gravitational field of Earth by analyzing the laser-ranged data of
some orbiting, or proposed, laser-ranged geodetic \sat s. Such a
measurement of the perigee advance would place limits on
hypothetical, very weak, Yukawa-type components of the
gravitational interaction with a finite range of the order of
$10^4$ km. Thus, we show that, at the present level of knowledge
of the orbital perturbations, the relative accuracy, achievable
with suitably combined orbital elements of LAGEOS and LAGEOS II,
is of the order of $10^{\rm -3}$. With the corresponding measured
value of $\ppn$, by using $\et=4\beta-\gamma-3$ from Lunar Laser
Ranging, we could get an estimate of the PPN parameters $\gamma$
and $\beta$ with an accuracy of the order of $10^{-2}-10^{-3}$.
Nevertheless, these accuracies would be substantially improved in
the near future with the new Earth gravity field models by the
CHAMP and GRACE missions. The use of the perigee of LARES (LAser
RElativity Satellite), with a suitable combination of orbital
residuals including also the node and the perigee of  LAGEOS II,
would also further improve the accuracy of the proposed
measurement.}

\end{titlepage} \newpage \pagestyle{myheadings} \setcounter{page}{1}
\vspace{0.2cm}
\baselineskip 14pt

\setcounter{footnote}{0}
\setlength{\baselineskip}{1.5\baselineskip}
\renewcommand{\theequation}{\mbox{$\arabic{equation}$}}
\noindent

\section{Introduction}
The well known secular shift of the pericenter
$\omega$ of a test body induced by the Schwarzschild metric generated
by a static, spherically symmetric distribution of
mass [{\it Misner et al.,} 1973; {\it Ciufolini and
Wheeler,} 1995] is one of the classical tests of
Einstein's General Relativity.
\subsection{The Solar System tests}
The advance of pericenter, which can be expressed in terms of
$\ppn$, where $\beta$ and $\gamma$ are the
Eddington--Robertson--Schiff PPN parameters [{\it Will}, 1993],
has been measured in the Solar System, on the Mercury perihelion,
by radar signals transmitted from Earth to Mercury and back to
Earth [{\it Shapiro et al.,} 1972], yielding $\ppn=1.005\pm
7\times 10^{\rm -3}$. The contribution of possible systematic
errors raises the uncertainty to about $2\times 10^{\rm -2}$.
 From the analysis of ten years of data the relative error
published in [{\it Shapiro et al.}, 1976] amounts to 5$\times
10^{-3}$.
The major sources of systematic error in this measurement are the
poorly-known variations in the topography of the planet's surface
and the uncertainties in the radar scattering law [{\it Shapiro,}
1990; {\it Pitjeva}, 1993]. For a review of periastron advance
measurements on binary pulsars see [{\it Will}, 2001] and
references therein.



Several space missions have been proposed in order to measure,
among other effects, this phenomenon by using a variety of techniques:
the most recent and promising are SORT, IPLR and ASTROD [{\it Ni},
2001].

More accurate measurements of $\ppn$ might be performed in the
future by means of the ESA BepiColombo Mercury orbiter [{\it
Balogh et al.,} 2000; {\it Milani et al.,} 2001a; {\it Milani et
al.}, 2001b]. In [{\it Milani et al.}, 2001b] it is claimed that
the BepiColombo mission, which should be launched in 2009, should
allow to measure $\gamma$ and $\beta$ with a relative accuracy of
the order of $2\times 10^{-6}$.
%
\subsection{The proposed experiment}
In this paper we propose to perform a measurement of the general
relativistic ``gravitoelectric" pericenter advance in the
gravitational field of the Earth by using some suitable
combinations of the orbital residuals of existing or proposed,
spherical, passive, geodetic laser-ranged satellites [{\it Iorio},
2002a] with particular emphasis to \lg\ and \lgg\, in order to
exploit the relevant experience obtained with the \grc\
Lense-Thirring measurements in [{\it Ciufolini et al.}, 1996;
1997; 1998]. In [{\it Ciufolini and Matzner,} 1992] a
determination of the LAGEOS general relativistic gravitoelectric
perigee shift in the field of the Earth is reported, but the
accuracy amounts only to $2\times 10^{\rm -1}$.

This paper is organized as follows: in section 2 we compare the
present experimental accuracy in satellite laser-ranging
measurements with the relativistic expression of the perigee
shift of the two \lg\ satellites. In section 3 we analyze some of
the most important sources of systematic errors. The accuracy
achievable in measuring $\beta$ and $\gamma$ is also discussed.
By detecting the relativistic perigee rate of the proposed LARES
[{\it Ciufolini}, 1998] laser-ranged satellite, in conjunction
with the node and the perigee of LAGEOS II, it might be possible
to further improve the accuracy in the measurement of these
parameters over a long enough time span. This is the subject of
section 4 together with an analysis of the potential role of the
existing laser-ranged satellite Starlette. Section 5 is devoted to
the conclusions.
\section{The relativistic perigee precession of LAGEOS type satellites}

As known, in the slow-motion and weak-field approximation, the
Schwarzschild metric generated by a static, spherically symmetric
distribution of mass-energy induces an additional post-Newtonian
``gravitoelectric force" which acts on the orbit of a test body by
shifting its pericenter; in the PPN formalism the pericenter rate
can be written as: \eqi \dot\og_{\rm GE}=\rp{3 n G M}{c^2 a
(1-e^2)}\times\ppn,\lb{peg}\eqf in which $G$ is the Newtonian
gravitational constant, $c$ is the speed of light in vacuum,
$M$ is the mass of the central object, $a$ and $e$ are
semimajor axis and eccentricity, respectively, of the orbit of
the test body and $n=\sqrt{GM/a^{\rm 3}}$ is its mean motion. In
the following we define $\n\equiv \ppn$. General Relativity
predicts that the perigee shifts for \lg\ and \lgg\ amount to
3,312.35 milliarcseconds per year (mas/y in the following) and
3,387.46 mas/y respectively.

The actual experimental precision allows for detecting such rates
for both the \lg\ satellites [{\it Ciufolini,} 1996]. Indeed, for
the perigee the observable quantity is $r \equiv ea\dot\og$; at
present the RMS of the residuals of the best orbital fits amounts
to $\d r_{\rm exp} \sim 1$ cm, or less, for the two LAGEOS
satellites, over several orbits and with a set of force models
(i.e. not including modeling errors). Since the \lg\ eccentricity
is $e_{\rm I}=4.5\times 10^{\rm -3}$ the accuracy in detecting the
perigee is $\d\og^{\rm I}_{\rm exp} \sim \d r_{\rm exp}/(e_{\rm
I}a_{\rm I}) \simeq 37$ mas. So that, over 1 year the relative
accuracy in the measurement of the relativistic perigee shift
would be of the order of $ 1 \times 10^{\rm -2}$. For \lgg\ this
measurement accuracy is better than for \lg. Indeed, the \lgg\
eccentricity is $e_{\rm II}=1.4\times 10^{\rm -2}$, thus
$\d\og^{\rm II}_{\rm exp} \sim 12$ mas; this may yield an accuracy
of the order of $3\times 10^{\rm -3}$ over 1 year. Over 8 years,
by assuming the same fit error as before, it would amount to $
\sim 4\times 10^{-4}$. These considerations rule out the
possibility of directly using the perigee of the other existing,
spherical, geodetic laser-ranged satellites Etalon-1, Etalon-2,
Ajisai, Stella, Westpac-1 because their eccentricities are even
smaller than that of \lg. On the contrary, Starlette has an
eccentricity of the order of $2\times 10^{\rm -2}$; however, since
it orbits at a lower altitude is more sensitive than the \lg\
satellites to atmospheric drag and to Earth's zonal harmonics, so
that would be difficult to process its data at an acceptable level
of accuracy (See also section 4). Accordingly, in order to detect
the gravitoelectric relativistic shift in the gravitational field
of Earth, we will focus on the perigee of \lg\ and especially of
\lgg\ .
\section{The systematic errors}
The perigee of an Earth satellite is a "non-clean" orbital
element in the sense that it is affected by a large number of
aliasing classical forces inducing systematic errors which must
be carefully identified and analyzed.
\subsection{The static geopotential error}
The most important source of systematic errors in such a
measurement is represented by the mismodeling of the classical
perigee precession induced by the errors in even zonal harmonics
of the Earth's gravitational field [{\it Kaula}, 1966]. This error
is really critical because the resulting aliasing trend cannot be
removed from the data and nothing can be done about it apart from
assessing the related error as reliably as possible.

By using the covariance matrix of the EGM96 Earth gravity model
[{\it Lemoine et al.,} 1998] and adding the correlated terms in a
root-sum-square fashion up to degree $l=20$, we obtain for \lg\ a
systematic error $\d\n/\n_{\rm zonals}=8.1\times 10^{\rm -3}$
whereas for \lgg\ we have $\d\n/\n_{\rm zonals}=1.5\times 10^{\rm
-2}$. Since the major source of error lies in the uncertainty of
the first two mismodeled even zonal harmonics $\d J_2$ and $\d
J_4$, in order to eliminate most of the static and dynamical even
zonal terms of the geopotential, following the method in [{\it
Ciufolini,} 1996] for the measurement of the \grc\ \lt\ effect, we
search for suitable combinations of the orbital residuals of the
existing laser-ranged satellites. In Tab.\ref{combi} we report the
most promising combinations: their general form is \eqi
\d\dot\og^{\rm II}+\somma{i=1}{N}c_{\rm i}\d\dot\O^{\rm i}+c_{\rm
N+1}\d\dot\og^{\rm I}=x_{\rm GR}\ \n,\lb{zione}\eqf in where $N$
is the number of the nodes of different laser-ranged satellites
employed, $x_{\rm GR}$ is the slope, in mas/y, of the
relativistic trend to be measured, in Tab.\ref{combi}
$\d\n/\n_{\rm zonals}$ is the systematic error induced by the
even zonal harmonics up to degree $l=20$ calculated with the
EGM96 covariance matrix. The coefficients $c_i$ are determined in
terms of the orbital parameters of the satellites entering the
combinations.

\begin{table}[ht!]
\caption{Combined residuals: numerical values}\label{combi}
\begin{center}\begin{tabular}{lllllll}\noalign{\hrule height 1.5pt}
& $\O^{\rm II}$ & $\O^{\rm I}$ & $\O^{\rm Aj}$ & $\og^{\rm
I}$\\
\hline $\k$ & $c_1$ & $c_2$ & $c_3$ & $c_4$ & $x_{\rm GR}$
(mas/y)
& $\d\n/\n_{\rm zonals}$
\\
\hline

{\sl 1} & $-0.87$ & $-2.86$ & 0 & 0 & 3,387.46 &
$6.59\times 10^{\rm -3}$ \\

{\sl 2} & $-2.51$ & $-4.37$ & 0 & 2.51 & 11,704.92 &
$1.1\times 10^{\rm -3}$\\

{\sl 3} & $-1.96$ & $-3.69$ & $0.037$ & $1.37$ & 7,928.51
&
$8.1\times
10^{\rm -4}$\\

\hline \noalign{\hrule height 1.5pt}
\end{tabular}
\end{center}
\end{table}

It is important to stress that the use of the LAGEOS satellites,
due to their altitude, makes our measurement substantially
insensitive to the errors in the zonal harmonics of degree $l>20$,
so that our estimates of $\d\n/\n_{\rm zonals}$ presented here are
valid even in the case that the EGM96 covariance matrix for
degrees higher than $l>20$ would not be accurate enough.

In addition to \lg\ and \lgg\ we have only considered Ajisai [{\it
Iorio}, 2002a] since it is well tracked, contrary for example to
the Etalon satellites, and it would be less demanding than for the
other satellites to reduce its laser-ranged data to a level of
accuracy comparable to that of \lg\ and \lgg. Moreover, the other
laser-ranged \sat s orbit at lower altitudes, therefore they are
more sensitive to terms of the geopotential of higher degree.
Consequently, as confirmed by numerical calculations, the
inclusion of their data in the combined residuals would increase
$\d\n/\n_{\rm zonals}$\footnote{About combination \textsl{3}, it
should be noted that, in order to obtain more reliable and
accurate estimates of the systematic error due to the even zonal
harmonics of the geopotential, the calculation of the error should
include terms with degree higher than $l=20$ due to the
sensitivity of Ajisai to such higher degree terms. }.

It is important to point out that the uncertainties quoted in
Tab.\ref{combi} for $\d\n/\n_{\rm zonals}$ will be reduced in the
near future when the new gravity models from the CHAMP and GRACE
missions will be released [{\it Pavlis}, 2002a].
\subsection{The time--dependent systematic errors}
In regard to the evaluation of the impact of the other sources of
systematic errors, the role of the coefficients $c_i$ entering
the combinations has to be taken into account, indeed, according
to their magnitude, they can reduce or emphasize the effects of
the perturbing forces acting on the orbital elements weighted by
them.

For example, for the combination \textsl{2} of Table 1, over 1
year, the impact of the error in measuring the perigee rate of
LAGEOS amounts to about $2.5\times(3.1\times 10^{\rm
-3})=7.7\times 10^{\rm -3}$ while for combination \textsl{3} it
is about $1.37\times(4.7\times 10^{\rm -3})=6.4\times 10^{\rm
-3}$.

In this paper we only consider in detail the combination {\sl 1}
of Table 1
  \eqi\d\dot\og^{\rm II}-0.87\times
\d\dot\O^{\rm II}-2.86\times\d\dot\O^{\rm I}=3,387.46\times\n
,\lb{rsd}\eqf so that we exploit the insight acquired with the \lt\
\lg\ experiment [{\it Ciufolini et al.,} 1996; 1997; 1998].

The long-period harmonic perturbations, according to their
periods $P$ and to the adopted observational time span $T_{\rm
obs}$, may turn out to be less insidious than the mismodeled
secular perturbations due to the zonal harmonics of the
geopotential since, if $P<T_{\rm obs}$ and $T_{\rm obs}=nP,\
n=1,2,...$ they average out; if their periods are shorter than the
time span they can be viewed as empirically fitted quantities
which can be subsequently removed from the signal.

The results and estimates recently obtained for the LAGEOS
gravitomagnetic experiment can be applied to our proposed
measurement. Indeed, the solid Earth and ocean tidal perturbations
on LAGEOS and LAGEOS II and their impact on the Lense--Thirring
effect have been studied in [{\it Iorio}, 2001; {\it Iorio and
Pavlis}, 2001; {\it Pavlis and Iorio}, 2002], while the role
played by the non--gravitational perturbations has been analyzed
in [{\it Lucchesi}, 2001; 2002]. Regarding the atmospheric drag,
whose impact might be thought of as a serious drawback, especially
for the perigee of LAGEOS II, in [{\it Ciufolini et al.}, 1997] it
is shown that it essentially averages out over an orbital
revolution.

In Tab.\ref{8anni} we summarize the results obtained with the
proposed combination using LAGEOS and LAGEOS II for a 8-year
time span with orbital fits, arcs, each of 7 days. In assessing the total
systematic error we have accounted for the fact that the
gravitational errors are not independent, indeed we have simply summed them
up; then we have added these uncertainties and the other independent
systematic errors in a root-sum-square fashion.
In Tab.\ref{8anni} we can observe that the error budget is
mainly dominated by the even zonal harmonics and by the LAGEOS II
perigee measurement error.
In regard to the non--gravitational errors, according to the
results of [{\it Lucchesi}, 2001] for the direct solar radiation
pressure and the Earth's albedo and to the results of [{\it
Lucchesi}, 2002] for the thermal thrust perturbations and the
asymmetric reflectivity, the corresponding uncertainty would
amount to almost $1\times 10^{-2}$ over 7 years. In obtaining this
result a very pessimistic approach has been adopted by assuming a
mismodeling of 20$\%$ for all the perturbing effects except for
the direct solar radiation pressure which has been assumed to have
an uncertainty at the 0.5$\%$ level. However, we stress that only
the Earth's thermal thrust, or Yarkovski-Rubincam effect, induces
a mismodeled linear trend whose impact would amount to about
$1\times 10^{-4}$: the other non--gravitational forces are
time--varying with known periodicities and can be fitted and
removed from the signal as done for the tidal perturbations.

\begin{table}[ht!]
\caption[PPN preliminar error budget: $T_{obs}=8$ years, $\D t=7$
days.]{Preliminary error budget: $T_{obs}=8$ years, $\D t=7$
days.}\label{8anni}
\begin{center}
\begin{tabular}{ll}\noalign{\hrule height 1.5pt}

Even zonal harmonics (par. 3.2) & $6.59\times 10^{-3}$  \\

$J_3$ geopotential [{\it Pavlis and Iorio}, 2002]& $3.2\times 10^{-4}$ \\

Tides [{\it Iorio}, 2001; {\it Pavlis and Iorio}, 2002] & $4.4\times
10^{-4}$ \\

Non-gravitational effects [{\it Lucchesi}, 2001; 2002] & 1$\times
10^{-4}$
\\


Measurement error in LAGEOS II perigee (sect. 2) & $4\times 10^{-4}$ \\

\hline

Total systematic error & $7.3\times 10^{-3}$\\



\hline \noalign{\hrule height 1.5pt}
\end{tabular}
\end{center}
\end{table}

Forthcoming improvements in the Earth gravity field modeling
achievable with the CHAMP and GRACE data have been recently
extensively studied and reported in [{\it Pavlis}, 2002a]. They will yield
important improvements in the accuracy of our experiment. We
finally observe that we already have the required laser-ranging
data since, at present, we have almost 10 years of LAGEOS and
LAGEOS II data.

\subsection{The parameters $\beta$ and $\gamma$}
The results obtained for the combination {\sl 1} examined here,
together with $\delta\eta=8\times 10^{-4}$ reported in [{\it
Anderson and Williams,} 2001] for the combination
$\et=4\beta-\gamma -3$ [{\it Nordtvedt,} 1968; 1991], would allow
to measure $\b=\rp{2}{7}\et+\rp{3}{7}\n+\rp{4}{7}$ at the level $
\sim 3\times 10^{\rm
-3}$ independently
of other measurements of $\ga$; this result, which is of the same
order of magnitude of that
obtained with radar ranging [{\it Shapiro}, 1990],
should be compared with the most recent $\d\b=4.7\times 10^{\rm
-4}$, obtained with LLR data [{\it Anderson and Williams,}
2001].

The parameter $\gamma$, that may be written as
$\gamma=\frac{1}{7}\eta+\frac{12}{7}\nu-\frac{5}{7}$, would be
measured less precisely: $\delta\gamma=1.2\times 10^{\rm -2}$.
However, $\gamma$ can be directly measured via light deflection or
radar time delay. For example, in [{\it Fr\oe schl{\'{e}} et al.,}
1997] the result $\ga=0.997\pm 3\times 10^{\rm -3}$ is based on
the astrometric observations of electromagnetic waves deflection
in the visible frequency. By using the Viking time delay [{\it
Reasenberg et al.,} 1979], $\ga=1.000\pm 2\times 10^{\rm -3}$ was
obtained; the quoted uncertainty allows for possible systematic
errors. A more recent measurement based on the time delay with
the NEAR spacecraft [{\it Elliott et al.}, 1998] claims an
uncertainty $\leq 10^{\rm -3}$.

For an updated overview of the present status of the measurements
of the PPN parameters, see [{\it Will}, 2001]

\section{The role of LARES and Starlette}
The proposed LARES satellite [{\it Ciufolini}, 1998] will enable us to
perform several general relativistic tests and geophysical
measurements. In this section we investigate its role for the
measurement of the gravitoelectric perigee advance that would be
well detectable on LARES thanks to its proposed relatively large
eccentricity: $e_{\rm LR}\simeq0.04$.

A possible observable is the following combination of
orbital residuals \eqi\delta\dot\omega^{\rm
II}+c_1\delta\dot\omega^{\rm LR}+c_2\delta\dot\Omega^{\rm
II}=x_{\rm GR}\nu,\lb{lares}\eqf with \bea
c_1 & = & -4.71,\\
c_2 & = & 2.26,\\
  x_{\rm GR} & = & -12,117.4\ {\rm mas/y}. \eea The coefficients
$c_i$ depend on the orbital parameters of the selected satellites
and have been calculated corresponding to the proposed orbital
parameters of LARES: $a_{\rm LR}=12,270$ km, $i_{\rm LR }= 70$
deg and $e_{\rm LR}=0.04$. Let us now calculate the systematic
errors induced by the even zonal harmonics of the geopotential,
according to the EGM96 gravity model. \Rfr{lares} would allow to
cancel out the errors due to the first four even zonal harmonics;
the impact of the remaining ones, with degree higher than four,
amounts to an uncertainty of about $6\times 10^{-3}$. This error
will further reduce when the new gravity models from the CHAMP
and GRACE missions will be available. An interesting feature of
\rfr{lares} is that the impact of the non--gravitational
perturbations would produce an uncertainty of about $3.9\times
10^{-3}$ over a time span of 7 years, according to the very
conservative estimates discussed above; in particular the Earth
thermal thrust, i.e. the so-called Yarkovski--Rubincam effect,
would induce a mismodeled secular trend with an uncertainty of
only $6.4\times 10^{-5}$. The inclusion of LARES would thus
represent an improvement with respect to the LAGEOS and LAGEOS II
scenario previously outlined, especially when, in the near future,
the impact of the mismodeled non--gravitational perturbations will
increase relatively to the gravitational perturbations thanks to
the more accurate Earth's gravity field models.

   Among the existing laser-ranged satellites, Starlette, with its
   relatively large eccentricity of 0.0204, seems to be a natural candidate
   for the measurement of the perigee advance. Unfortunately,
   it orbits at an altitude much lower than the LAGEOS
   satellites ($a_{\rm Str}=7,331$ km) and thus the mismodeling of the
even zonal harmonics
   of the geopotential would induce large systematic errors in
   the measurement of the perigee rate. It turns out that the inclusion of the
   perigee of Starlette in the combined residuals if, on the one hand,
provides a
   further observable and thus would allow to
   cancel out the error of an additional mismodeled even zonal harmonic,
   on the other hand it raises the total error due to
   the geopotential, even with the LARES data.
   Indeed, in a combination including the nodes of LAGEOS and LAGEOS
   II and the perigees of LAGEOS II and Starlette, the relative error
   due to the geopotential amounts to only $6.2\times 10^{-2}$. This
rules out the possibility
   of using the Starlette data, at present level of knowledge of the Earth's gravitational field,
   to improve the measurement
of the relativistic perigee  advance.

\section{Conclusions} The main features of our proposed
measurement of perigee advance using laser-ranged satellites are
the following:
\begin{itemize}
   \item It will provide a test of a general
   relativistic effect, i.e. the pericenter advance, {\it in the gravitational
   field of Earth}, complementary to similar tests in the field of Sun.
   Indeed, such a measurement in the gravitational
   field of Earth would place stronger limits
   on possible, very weak, Yukawa-type components of the
   gravitational interaction with a finite range of the order of $10^4$ km [{\it Iorio}, 2002b].
   \item For a suitable combination of the orbital residuals of the
   nodes of LAGEOS and LAGEOS II and perigee of LAGEOS II,
   the relative error due to the mismodeled classical precessions by
the static even zonal
   harmonics of the geopotential,
   which is the most insidious source of systematic
   error, amounts, according to the EGM96 Earth gravity
   model, to about $6\times 10^{-3}$. The total error in our proposed
combination of orbital
   elements, also due to the other time--dependent systematic errors,
   amounts, over an observational time span of eight years, to almost $7\times
   10^{-3}$.
   \item In the near  future the knowledge of the Earth gravitational
field will increase   thanks to the
   data from the CHAMP and GRACE missions. Moreover, the rms accuracy
   in measuring the position of the LAGEOS satellites should reach the few
   mm level. Consequently, there will be an increase in the
   accuracy of our measurement.
   \item  In regard to the terrestrial space
    environment, which induces a large number of orbital perturbations
  on the satellites,
  it is very well studied and the action of these perturbations,
    especially on the LAGEOS satellites, is very well modeled. The
data processing would be reliably based on
   the large experience collected in several years of analysis of
   the LAGEOS orbits [{\it Pavlis}, 2002b].
    Moreover,
  it would be relatively fast and easy to include the future
improvements in the force modeling
    in our analysis, e.g. the improvements due to a better
experimental and theoretical
    knowledge
    of the rotation rate and spin axis orientation of LAGEOS type satellites.
    \item The proposed measurement, as an additional outcome, from $\ppn$ in
    conjunction with $\et=4\beta-\gamma-3$ from Lunar Laser Ranging,
would allow to obtain
    the PPN
    parameters $\gamma$ and $\beta$ in the gravitational field of
Earth (see above) with an accuracy at the level of $10^{-2}-10^{-3}$.
    This accuracy would  improve with
    the new Earth's gravitational models from CHAMP and GRACE.
    \item Starlette data, although this satellite has a relatively
eccentric orbit, should not
   be included in the combined residuals. Indeed, Starlette
   would raise the systematic error in the geopotential due to its
lower altitude with respect to the
   LAGEOS satellites
   and thus due to its higher sensitivity to the even zonal
   harmonics of the terrestrial gravitational field.
   \item The use of the perigee of the proposed LARES satellite, for a
suitable combination
   of observables including also the node
   and the perigee of LAGEOS II, would
   allow to reduce the systematic relative error due to the mismodeled even
   zonal harmonics of the geopotential to about 6$ \times 10^{-3}$,
according to EGM96.
   This proposed measurement would also take advantage of the new
Earth gravity models from CHAMP and GRACE.
   According to conservative
   estimates, the impact of the non--gravitational perturbations would
amount to about $4\times 10^{-3}$
   over 7 years; in particular the impact of the mismodeled Earth
thermal thrust, or Yarkovski-Rubincam
   effect,
   would amount to only about $6\times 10^{-5}$.
   \end{itemize}
\section*{Acknowledgements}L. I. would like to thank L. Guerriero for
his encouragement and support, C.M. Will and I.I. Shapiro for the
useful information kindly supplied to him.
E. C. Pavlis gratefully acknowledges NIMA support through the NURI
grant NMA201-01-1-2008.

\end{document}